\documentclass[final,english]{bullsrsl}

\usepackage[latin1]{inputenc}

\usepackage[T1]{fontenc}

\usepackage{natbib} 

\usepackage{graphicx}
\usepackage{xcolor}

\begin{document}
\title{Investigating Variability in Long-lived Protoplanetary Disks (>10 Myr) using NEOWISE}



  \author[affil={1}, corresponding]{Gregory Mathews}{Ben}
  \author[affil={1}, corresponding]{Jessy}{Jose}
  \author[affil={2}]{Jes\'us}{Hern\'andez}

\affiliation[1]{Department of Physics, Indian Institute of Science Education and Research Tirupati, Yerpedu, Tirupati - 517619, Andhra Pradesh, India}
\affiliation[2]{Instituto de Astronom\'ia, Universidad Nacional Aut\'onoma de M\'exico, Apartado Postal 106, C. P. 22800, Ensenada, B. C., M\'exico}


\correspondance{jessyvjose1@gmail.com, gregorygmb10@gmail.com}

\maketitle
\begin{abstract}
We present the mid-infrared variability study of 120 long-lived disk candidates (ages 10-100 Myr) and 13 confirmed "Peter Pan" disks using  NEOWISE to constrain the longevity of dynamic disk processes beyond the standard 10 Myr dispersal threshold. We identified 32 sources exhibiting significant variability by combining stochastic flux metrics and the Stetson $J$ index. The majority of these variables 
belong 
to 
the 
10-25 Myr age range, suggesting a sharp decline in disk brightness variability after 25 Myr. Morphologically, the variability is characterized by irregular fluctuations with mean amplitudes of $\Delta W2 \approx 0.5$ mag, consistent with variable extinction or inner disk warping induced by embedded planetesimals. We also report a strong correspondence between mid-infrared and optical variability for the Peter Pan disks, with all 9 mid-IR variables confirmed as optical variables in the literature. Our findings reveal a population of 23 new long-lived disk candidates that retain active, gas-rich environments, supporting the existence of a rarefied class of systems that survive well into the debris disk era.
\end{abstract}





\section{Introduction}
Variability is a unique feature of Young 
Stellar
Objects (YSOs) and acts as an effective diagnostic for the physical mechanisms that regulate the early stages of stellar evolution. These photometric variations originate from a wide variety of causes, each of which dominates various parts of the star-disk system \citep{1945ApJ...102..168J,1994AJ....108.1906H}. Short-term variability usually comes from stellar rotation through cool magnetic spots or hot spots caused by accretion shocks on the stellar surface \citep{2015A&A...581A..66V,2017ApJ...836...41C}. On longer timeframes, variability is usually driven by structural changes in the circumstellar disk, such as varying extinction from warping or clumping, or by unstable accretion rates that cause strong bursts. \citep{2015AJ....150...32R,2017MNRAS.465.3011C,2018ApJ...854...31J,2025ApJ...987...23C}.

To fully characterize the star-disk interaction, time-domain studies must extend into the mid-infrared (MIR), where dust re-emission provides a direct view of disk geometry and heating \citep{2020MNRAS.499.1805L,2023ApJ...957....8W}. The Near-Earth Object Wide-field Infrared Survey Explorer (NEOWISE) provides a 11 year long-baseline window into this regime, allowing us to differentiate between variable extinction and accretion-driven luminosity changes. The survey has found several high-amplitude accretion outbursts, including rare FU Orionis and EX Lupi-type events that were optically entirely hidden, by monitoring a large sample of YSOs \citep{2018ApJ...869..146H,2021ApJ...920..132P,2023ApJS..264...38M,2023JKAS...56..253C,2024MNRAS.52711651A,2024ApJS..272...44L,2025ApJ...987...23C}.

Although mid-IR variability has been extensively investigated for young systems ($< 10$ Myr) where the disk retains its primordial gas reservoir, the variability properties of disks beyond this age remain largely unexplored. Standard models of disk evolution imply that primordial gas normally dissipates within 3-10 Myr \citep{2001ApJ...553L.153H,2018MNRAS.477.5191R,2024ApJ...970...88P}; however, recent studies have challenged this timescale. New evidence indicates that disks can persist for significantly longer durations, particularly around M-dwarf stars \citep{2022ApJ...939L..10P,2024ApJ...963..122P, 2025MNRAS.541.2246B,2025ApJ...994..248F,2025arXiv251206873P}. This is clearly illustrated by the discovery of "Peter Pan" discs (PPDs); gas-rich systems that continue to accrete well beyond 20 Myr, and in some cases up to $\sim$45 Myr \citep{2018MNRAS.476.3290M,Silverberg_2020,2025AJ....169..141W}. The existence of these survivors implies that the window for planet formation and migration may remain open far longer than previously thought. Yet, it remains unclear whether the accretion flows in these older systems exhibit the same stochastic volatility seen in their younger counterparts. In this work, we utilize the long-baseline NEOWISE archive to examine the mid-IR variability of a sample of these rare, long-lived disks to determine if they remain dynamic or
settle
into a steady state at longer ages.

\section{Targets and Datasets}
In this study, we investigate mid-infrared variability in a sample of 133 long-lived circumstellar disk systems with ages $>10$ Myr. Our sample comprises two components: (1) 120 newly identified long-lived disk candidates from a recent blind survey of nearby young clusters, exhibiting infrared excesses consistent with primordial "full" disks, and (2) 13 spectroscopically confirmed accreting PPDs from the literature, which serve as a control group for known accretion-driven variability. Below, we describe the target selection criteria (Section \ref{sec:target}) and the NEOWISE dataset used for our variability analysis (Section \ref{sec:neowise}).
\subsection{Target Selection}
\label{sec:target}
Our study focuses on identifying variability in long-lived disk candidates. To achieve this, we compiled a sample from two distinct sources:

\textbf{1. The Old Disk Survey ( Sample 1)}: The primary sample consists of 120 long-lived disk candidates selected from our recent blind survey of young clusters in the solar vicinity \citep{2025MNRAS.541.2246B}. These sources have estimated ages of 10-100 Myr and exhibit infrared excesses consistent with "full" primordial disks. Their membership was established via Gaia 
DR2 
astrometry by \citet{2020A&A...640A...1C} and ages were determined through isochronal analysis using PARSEC models. These sources were classified as 'full' disks based on their position in the $K-W3$ vs $K-W4$ color-color diagram \citep{2025MNRAS.541.2246B} , exhibiting distinct infrared excesses that separate them from transitional or debris disk systems. While this suggests the presence of substantial circumstellar material, their current accretion status remains unconfirmed spectroscopically.

\textbf{2. Known PPDs (Sample 2)} : To provide a comparative baseline for identifying accretion-driven variability, we included 13 confirmed PPDs from the literature \citep{Silverberg_2020,2025AJ....169..141W}. These long-lived discs ($>10$ Myr) were kinematically confirmed as moving group members by \citet{Silverberg_2020} or identified via isochronal analysis by \citet{2025AJ....169..141W}, with their active accretion status established through the spectroscopic detection of strong H$\alpha$ emission in both studies. Unlike Sample A, these objects are confirmed to be actively accreting, serving as a control group for what genuine, long-lived accretion activity looks like in the NEOWISE bands.

\subsection{NEOWISE Time-Domain Photometry}
\label{sec:neowise}
To characterize the mid-infrared variability of these targets, we utilize multi-epoch photometry from the NEOWISE \citep{2014ApJ...792...30M}. Following the reactivation of the WISE spacecraft in December 2013, the mission has surveyed the entire sky approximately once every six months at 3.4 $\mu$m ($W1$) and 4.6 $\mu$m ($W2$). This survey provides a broad time-domain catalogue with a baseline of about 11 years. The simultaneous observation of $W1$ and $W2$ data is particularly valuable for differentiating between variations in disk temperature and changes in extinction. We cross-matched the coordinates of both sample lists with the NEOWISE Single Exposure Source Table using a search radius of $1.0''$ ,successfully identifying NEOWISE counterparts for all 133 sources in our sample.

\section{Identifying Variable Sources}
We filtered the raw NEOWISE light curves to ensure the reliability of our variability analysis. First, we retained only those targets detected in more than five separate epochs in the $W2$ band to guarantee sufficient temporal coverage. Second, to minimise contamination from nearby sources, we required the source position to be stable, rejecting targets where the standard deviation of the radial distance from the catalog coordinates exceeded $0.3''$. This cut-off effectively filters out faint or saturated sources that commonly show large positional dispersions \citep{2021ApJ...920..132P}, ensuring we retain only reliably measured sources. Finally, we removed noisy data by selecting only those sources with a mean $W2$ measurement uncertainty smaller than 0.2 mag. All 133 sources in our sample satisfy all three selection criteria described above.  With this high-quality dataset established, we assessed the variability status of our sample of 133 sources using two complementary statistical methods: the classification scheme described by \citet{2021ApJ...920..132P} and the Stetson variability index\citep{1996PASP..108..851S}.
\subsection{Variability from Light Curve Statistics}
\label{sec:park}
Following the method of \citet{2021ApJ...920..132P}, we first combined the individual NEOWISE exposures into single data points for each six-month epoch. To remove outliers, we used only the middle 70\% of measurements in each epoch, discarding the top and bottom 15\%. The representative magnitude and date for each epoch were calculated as the average of this remaining data. The measurement error is calculated by adding, in quadrature, the mean error and the standard deviation (in magnitudes) of the exposures in each epoch. 

\begin{figure*}[t]
\centering
\includegraphics[width=\textwidth]{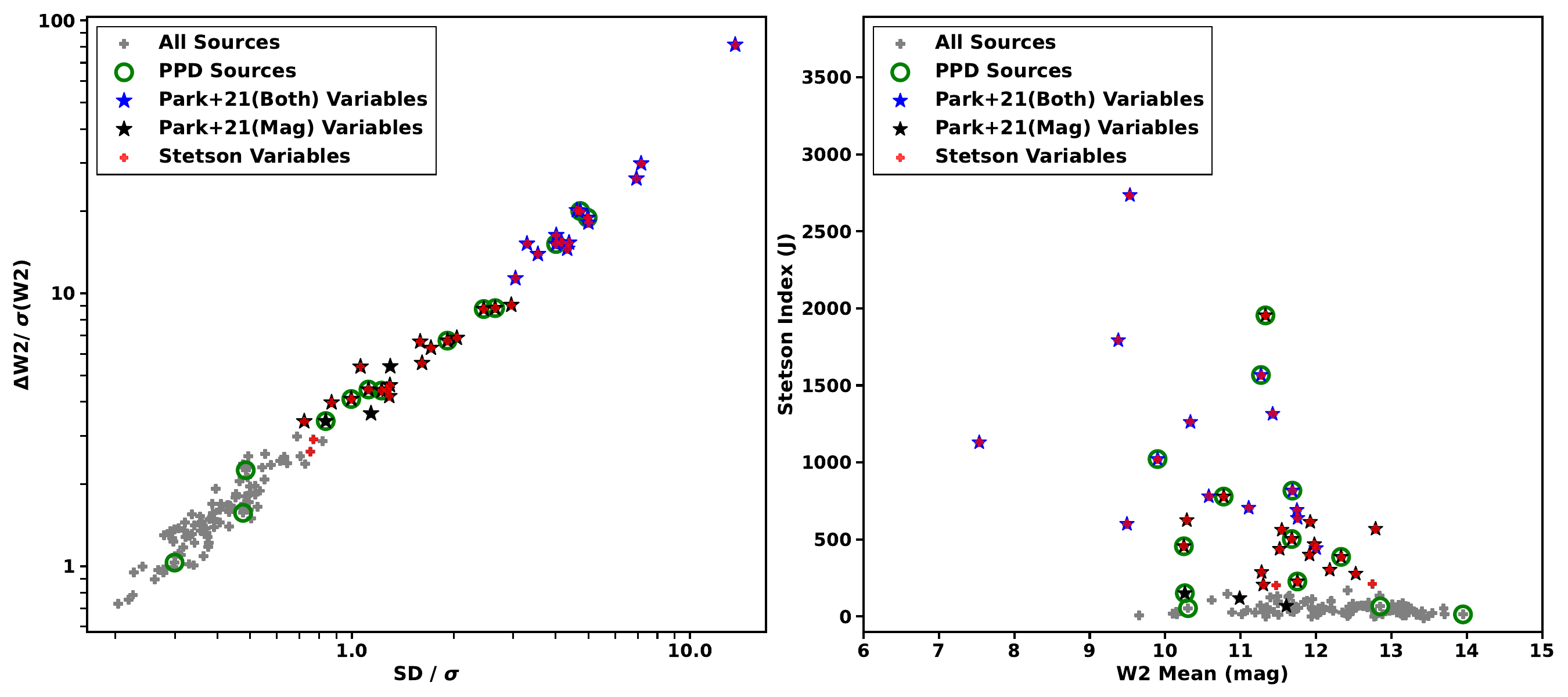}
\begin{minipage}{12cm}  
\centering
\caption{Variability analysis results. Left panel:The variability metric plot showing the standard deviation ($SD/\sigma$) versus the magnitude range ($\Delta W2/\sigma(W2)$). Right panel: The Stetson $J$ index versus mean $W2$ magnitude. The green circles indicate PPD sources in our sample. The red crosses show variable candidates selected by the Stetson criteria. Blue stars represent sources satisfying both \citet{2021ApJ...920..132P}criteria (stochastic and magnitude range), while black stars indicate sources satisfying only the magnitude range criterion.}
\label{fig:var}
\end{minipage}
\end{figure*}

To find variable sources, we examined the light curves in flux space. We initially converted the $W2$ magnitudes to flux and calculated the standard deviation over the whole light curve. We then divided this standard deviation by the mean flux uncertainty ($\sigma$),which is the average of the individual epoch errors stated earlier. Sources were categorized as variable if their ratio surpassed that of the noise 
\citep[$SD/\sigma > 3$;][]{2021ApJ...920..132P}. We also searched the complete magnitude range to capture sources with  slow secular trends or rare events that may have no effect on the global standard deviation. We regarded a source variable if the difference between its brightest and faintest epochs exceeded three times the average uncertainty ($\Delta W2 / \sigma(W2) > 3$) \citep{2021ApJ...920..132P}. Together, these conditions make sure that we identify both high-amplitude brightness changes and continuous stochastic variations.
Applying these criteria to our primary sample of 120 long-lived disk candidates (Sample 1), we identified 25 sources exhibiting significant mid-infrared variability (Figure \ref{fig:var}, left panel). In the figure, sources satisfying both the stochastic ($SD/\sigma > 3$) and magnitude range ($\Delta W2 / \sigma > 3$) criteria are represented by blue stars (13 sources), indicating robust variability. Those satisfying only the magnitude range criterion are shown as black stars (12 sources). For our control group of 13 confirmed PPDs (Sample 2), marked with green circles, 10 sources were identified as variable, with 3 satisfying both checks and 7 satisfying only the magnitude range criterion.

\subsection{Variability from Stetson Index(J)}
\label{sec:stet}
As an additional statistic, we used the Stetson $J$ index, which uses correlated signals to discriminate between physical variability and random noise. The $J$ index, in contrast to basic deviation statistics, weights pairs of simultaneous observations ($W1$ and $W2$ bands in 
the
case of NEOWISE). The index is based on the idea that actual astrophysical variability will cause correlated brightness variations across many bands (or repeated observations within a short window), resulting in a positive contribution to the index, whereas uncorrelated noise will average to zero. A high $J$ value serves as a strong indicator of inherent variability \citep{2014AJ....147...82C,2025MNRAS.541.1866C}, especially in the case of low-amplitude signals that would be minimally identified by single-band flux statistics.

Figure \ref{fig:var} right-panel shows the J value vs mean $W2$ magnitude distribution of all the sources. Variable candidates separate clearly from the non-variable population, which clusters near $J \approx 0$. To formally identify these variables, we applied a selection threshold based on the sample statistics: sources with a Stetson index greater than three times the median (i.e., $J > 171$) were classified as variables. Using this criterion, we identified 25 variables in Sample 1 and 9 variables in Sample 2.
\subsection{Final Variable Sample}

To create a high-confidence catalog of variable long-lived disks, we selected only those sources that were independently identified by both the Light curve statistics (Section \ref{sec:park}) and the Stetson $J$ index (Section \ref{sec:stet}). This cross-validation eliminates any false positives caused by transitory noise or single-band artifacts.  This intersection yielded a final sample of 32 robust variables out of 133 sources ($\approx24\%$), comprising 23 long-lived disk candidates (Sample 1) and 9 confirmed PPDs (Sample 2). These 23  candidates represent a population of "old" ($>10$ Myr) disks that exhibit mid-IR photometric variability.
\section{Discussion}
Our variability analysis has identified 32 robust mid-infrared variables among 133 long-lived disk systems ($\approx$ 24\%), comprising both newly identified disk candidates and known PPDs. These detections demonstrate that a significant fraction of evolved circumstellar disks exhibit significant mid-infrared variability well beyond the typical dispersal timescale. In this section, we interpret these findings by examining the age distributions of the variables, analyzing their amplitude and morphology, and comparing them with known PPDs.
\subsection{Evolutionary Status of Variable Sources}
The duration of dynamic disk variability is critical to our knowledge of planet formation, with most primordial disks expected to disperse within 3-10 Myr. To constrain the timescales over which disks remain dynamically active, we analyzed the distribution of variability strength as a function of system age. Figure \ref{fig:Jage} presents the Stetson $J$ index versus estimated cluster age for our full catalog of variable sources (Sample 1 and Sample 2 combined). The distribution shows that 
the
majority of variables (blue crosses) belong to the 10-25 Myr age range,  with 25 out of 32 sources (78\%) younger than 25 Myr and the remaining 7 (22\%) falling in the 25-50 Myr range. Although the actual disk structure may endure up to $\sim$100, this clustering indicates that the dynamic processes that cause high-amplitude mid-IR fluctuations (such as vigorous accretion or turbulent disk warping) dramatically diminish after 20 Myr. 

\begin{figure}[h!]
\centering
\includegraphics[width=0.5\textwidth]{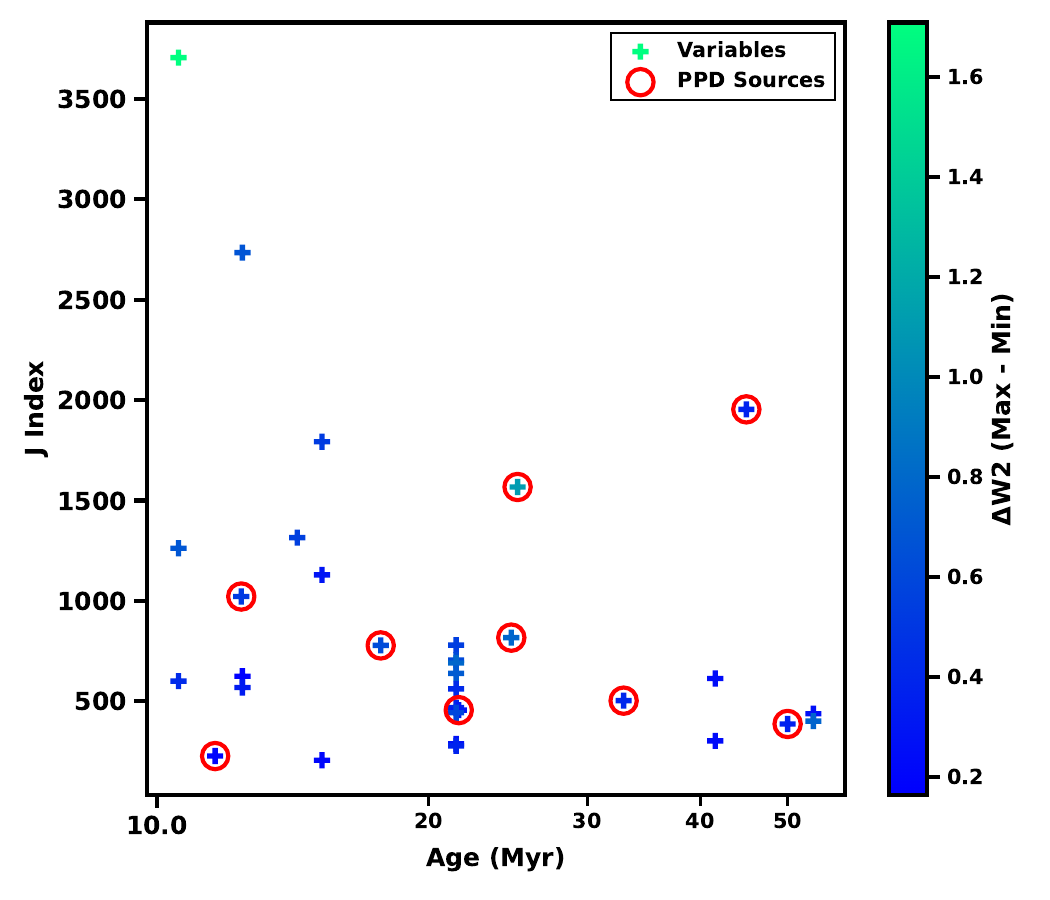}
\begin{minipage}{12cm}  
\centering
\caption{The plot shows the Stetson $J$ index versus cluster age for the identified variables (crosses). The color scale indicates the peak-to-peak magnitude range ($\Delta W2$). The known PPDs are highlighted with red circles.}
\label{fig:dw2}
\end{minipage}
\end{figure}

The distribution of the PPD candidates (red circles) further highlights the complexity of this evolution. While a majority of these known accreting systems are detected as variables (9/13), their behaviour is far from uniform. As seen in the plot, only two PPDs exhibit strong variability ($J > 1500$), while the remaining 
display
more moderate activity levels. This demonstrates that "Peter Pan" status does not ensure high-amplitude mid-IR variability at all epochs. However, the presence of active PPDs alongside our new candidates in the  25-50 Myr region supports the existence of a rare population of long-lived systems that survive the standard dispersal phase. We note that while variability provides a strong diagnostic, the definitive accretion status of these new candidates is yet to be confirmed via spectroscopic follow-up.

\subsection{Morphology of the Variable Sources}
\begin{figure}[h!]
\centering
\includegraphics[width=0.5\textwidth]{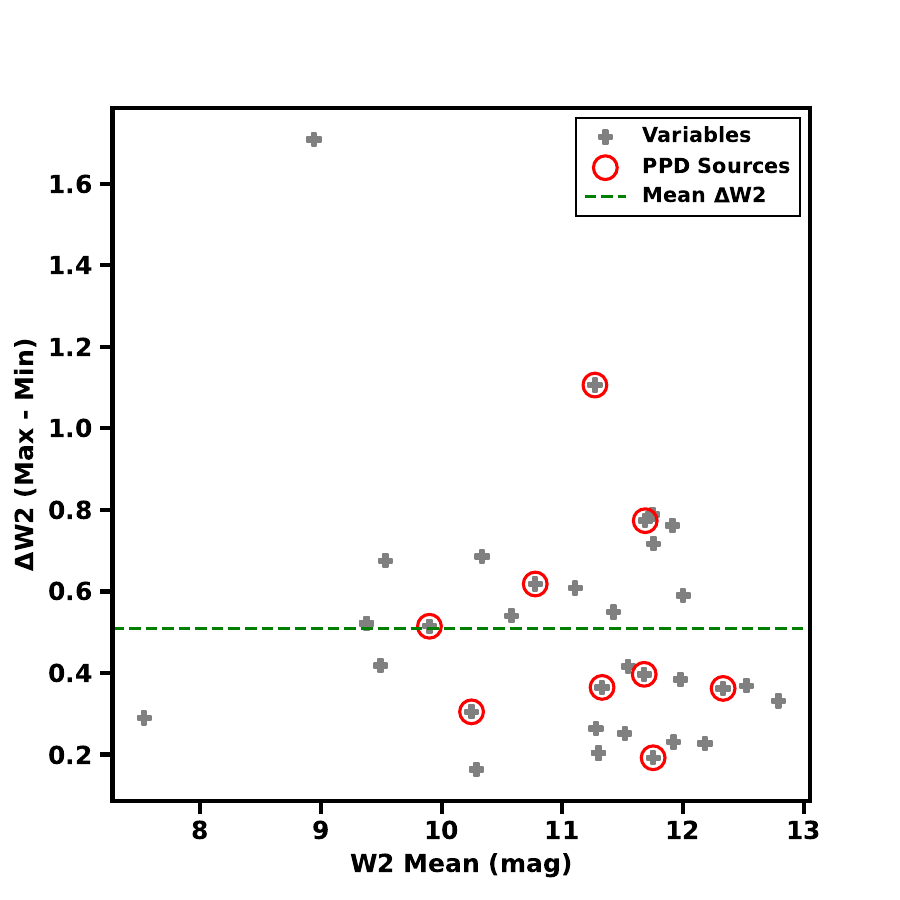}
\begin{minipage}{12cm}  
\centering
\caption{The plot shows the ($\Delta W2$) versus mean $W2$ distribution of the full sample of variable sources( grey crosses). The known PPD candidates are highlighted with red circles. The dashed green line indicates the sample median ($\Delta W2=0.51$ mag)}
\label{fig:Jage}
\end{minipage}
\end{figure}
We also explored the physical nature of the variation by determining the peak-to-peak amplitude range ($\Delta W2$) as a function of mean magnitude brightness (Figure \ref{fig:Jage}). The variable sources have $\Delta W2$ ranging from 0.2 to 1.6 mag and are concentrated around the mean amplitude of $\Delta W2=0.51$ mag, as shown by the dashed green line. The PPD sources (red circles) also follow the same trend, occupying the same amplitude range as Sample 1. In evolved systems, these irregular $\sim$0.5 mag fluctuations are consistent with variable extinction events or changes in the inner disk geometry \citep{2014AJ....147...82C,Silverberg_2020}. As predicted by current models, such structural irregularities could be caused by the dynamic coupling between the remaining disk material and the embedded planetesimals, which might lead to the warping of the inner disk or the formation of dusty clumps that periodically occult the central star\citep{2017MNRAS.470..202B,2018MNRAS.481...20N}.

Although the majority of the population shows this irregular variation, there are also individual objects with amplitudes well above the mean value ($\Delta W2 > 1.0$ mag). These objects could represent "burst" variables, possibly fueled by irregular magnetospheric accretion instabilities. If confirmed, this would suggest that a fraction of these evolved sources retain active gas reservoirs well beyond the standard dispersal era. Generally, variability caused by changing extinction can be distinguished from intrinsic luminosity variations (like accretion bursts) using multi-wavelength photometry. Extinction events are strongly wavelength-dependent, producing deeper flux dips in the optical compared to the mid-infrared. In contrast, intrinsic accretion bursts drive correlated increases in brightness across both optical and infrared bands. To further investigate these physical drivers, we plan to explore the optical variability of these sources using data from TESS and ZTF.

\subsection{Variability in PPDs}
The PPDs of Sample 2 offer key properties for benchmarking variability. The previous optical monitoring of these systems by \citet{Silverberg_2020} and \citet{2025AJ....169..141W} showed that they indeed display variability due to accretion and extinction. We found that all PPDs identified as mid-infrared variables in our study are also known optical variables, lending support to a common physical mechanism. The prototype system J0808 is one such example. Although \citet{Silverberg_2020} labeled it as an "aperiodic burster" and "dipper" based on accretion and dust extinction, our NEOWISE data analysis independently verifies it as a high-confidence variable with a high Stetson index ($J=1954$). This confirms that the accretion and extinction events observed in the optical are energetic enough to drive observable bulk heating changes in the inner disk dust.

One exception to this trend is J0501. Although \cite{Silverberg_2020} verified this source as an optical variable, it has a stable mid-infrared light curve in our analysis. Their interpretation of TESS data suggests that the optical variability is due to fixed starspots with a rotation period of 0.906 days. This is likely due to the fact that the contrast of the starspots decreases substantially at longer wavelengths, and the short rotation period is not well-sampled in the mid-infrared survey.  The stability of the mid-infrared light curve confirms that the optical variability is of photospheric origin, and not disk-related.  This example demonstrates that mid-infrared selection criteria effectively isolate structural or accretion-driven disk variability from stellar photospheric variability.
\section{Conclusion}
We investigated the mid-infrared variability of 133 old disk (10-100 Myr) sources (120 new candidates and 13 confirmed PPDs) using the 11-year baseline of the NEOWISE survey. We identified a robust sample of 32 variables ($\approx$24\% of our 133 sources). Our analysis revealed the following:
\begin{enumerate}
\item The majority of the identified variables (78\%) 
belong
to the 10-25 Myr age range, and the dynamic processes driving high-amplitude variability typically decline after 25 Myr. However, we have also identified 7 long-lived active sources, both new and existing PPDs, that remain active in the age range of 25-50 Myr, thus confirming that there exist rare cases where systems can maintain dynamic gas reservoirs beyond the standard dispersal ages.
\item Most sources exhibit irregular variability with mean amplitudes of $\Delta W2 \approx 0.5$ mag. These fluctuations are consistent with variable extinction or disk geometry changes, potentially induced by embedded planetesimals.
\item We found that the 9 PPDs identified as mid-infrared variables in this work are also confirmed optical variables. The non-detection of J0501 is likely because its starspot-driven optical variability is too weak to be detected in the mid-infrared.
\end{enumerate}



\begin{acknowledgments}
This publication makes use of data products from the Near-Earth Object Wide-field Infrared Survey Explorer (NEOWISE), which is a joint project of the Jet Propulsion Laboratory / California Institute of Technology and the University of California, Los Angeles. NEOWISE is funded by the National Aeronautics and Space Administration.
\end{acknowledgments}

\begin{furtherinformation}

\begin{orcids}

  \orcid{0009-0005-5716-3956}{Gregory Mathews}{Ben}
  \orcid{0000-0003-4908-4404}{Jessy}{Jose}
\end{orcids}


\begin{conflictsofinterest}
The authors declare that there is no conflict of interest.
\end{conflictsofinterest}

\end{furtherinformation}



%

\bibliographystyle{bullsrsl-en.bst}

\bibliography{extra.bib}

\end{document}